\documentclass[sn-mathphys-num]{sn-jnl}

\usepackage{graphicx}%
\usepackage{multirow}%
\usepackage{amsmath,amssymb,amsfonts}%
\usepackage{amsthm}%
\usepackage{mathrsfs}%
\usepackage[title]{appendix}%
\usepackage{xcolor}%
\usepackage{textcomp}%
\usepackage{manyfoot}%
\usepackage{booktabs}%
\usepackage{algorithm}%
\usepackage{algorithmicx}%
\usepackage{algpseudocode}%
\usepackage{listings}%

\theoremstyle{thmstyleone}%
\theoremstyle{thmstyletwo}%

\theoremstyle{thmstylethree}%

\begin{document}

\title[Half Landau-Zener ramp to a quantum phase transition in a dissipative single spin sodel]{Half Landau-Zener ramp to a quantum phase transition in a dissipative single spin model}


\author*[1]{\fnm{Sei} \sur{Suzuki}}\email{sei01@saitama-med.ac.jp}

\affil*[1]{\orgdiv{Department of Liberal Arts}, \orgname{Saitama Medical University}, \orgaddress{\city{Moroyama}, \postcode{350-0495}, \state{Saitama}, \country{Japan}}}


\abstract{We study the dynamics of a single spin coupled to a bosonic bath at zero temperature driven by a ramp of the bias field. A single spin coupled to a bosonic sub-Ohmic bath exhibits a quantum phase transition at a certain strength of spin-boson coupling. When the bias field is ramped from a large value to zero at this critical coupling strength, the system initialized at the ground state ends up with a finite magnetization due to the critical slowing down near the transition. On the basis of the pulse-impulse approximation, we derive a scaling law between the residual magnetization and the ramp speed. The obtained scaling relation is examined using a numerical simulation based on the tensor network. The data are in favor of the scaling law to hold. We discuss the demonstration of our theoretical results by means of quantum simulation using the quantum annealer.
}

\keywords{dissipative spin model, quantum phase transition, Kibble-Zurek scaling}



\maketitle

\section{Introduction}\label{sec1}

The quantum two-level system is not only the most fundamental model in the study of quantum mechanical dynamics but playing basic roles as a single qubit in quantum information and computation. When a bias of two levels is ramped with time so as to interchange the energy of two basis states, the state initialized at one of the two basis states ends up in a superposition of them. The transition from the initial energy level to the different level at the end is called as the Landau-Zener transition and its application ranges from atomic collision\cite{landau1977}, chemical reaction\cite{hanggi1990}, to macroscopic tunneling\cite{izmalkov2004}. In realistic situations, however, most systems cannot be free from its environment, hence it is significant to study the influence of an environment. The basic model for such a quantum two-level system coupled to an environment is the dissipative spin model\cite{leggett1987,weiss2012}. As far as the Ohmic boson bath is concerned, the dissipative spin model reduces to the anisotropic Kondo model. The dissipative spin model has attracted a long-lasting attention for more than three decades. It is known that a second-order quantum phase transition (QPT) takes place at a critical coupling strength between the spin and a bath, as far as a sub-Ohmic bath is concerned\cite{bulla2003}. Surprisingly, the probability of the Landau-Zener transition for the ramp of the bias field from an infinitely positive to infinitely negative values is not altered in the presence of the system-bath coupling\cite{wubs2006}, even though the coupling strength is fixed at a critical value and the ramp intersects a QPT. However, this is not true when the ramp is stopped halfway. If particular, when the ramp stops at a quantum critical point, the picture of two levels is no longer valid and a QPT plays a crucial role in the vicinity of a quantum critical point.

When a parameter determining the equilibrium state of a system is ramped linearly with time near a phase transition, a system acquires topological defects with density scaled by the speed of the ramp. This phenomenon, dubbed as the Kibble-Zurek mechanism (KZM), has been studied in broad context ranging from the cosmology to the condensed matter physics\cite{kibble1976-1,zurek1985-1}. A lot of attentions have been paid to KZM of several QPTs in the lase two decades\cite{polkovnikov2005-1,zurek2005-1,dziarmaga2005,dziarmaga2010,polkovnikov2011,dutta2015}. It has been known that, when a parameter governing the quantum fluctuation is ramped with the speed $v$ and stopped at an ordinary second-order QPT point, the defect density $n$ is scaled as $n\sim v^{d\nu/(z\nu + 1)}$ with $v$, where $d$ is the dimension of a system, $\nu$ is the correlation length critical exponent, and $z$ is the dynamical critical exponent. This so-called Kibble-Zurek scaling has been extended to a quantum critical line\cite{divakaran2008}, multicritical point\cite{mukherjee2007,deng2009,mukherjee2010}, anisotropic critical point\cite{sengupta2008,mukherjee2011,hikichi2010}, discontinuous critical point\cite{suzuki2015}, and nonlinear ramps\cite{sen2008,barankov2008-1}. Besides, the Kibble-Zurek scaling of a QPT has been tested by quantum simulations with an optical interferometer\cite{xu2014},  superconducting qubits\cite{gong2016}, trapped ions\cite{ulm2013,cui2016}, Rydberg atoms\cite{keesling2019}, and the quantum annealer\cite{gardas2018,bando2020,king2022,king2023}.

In the present work, we discuss KZM in the dissipative spin model with a ramp of the bias field. Since our model is defined in zero dimension, the defect density is not defined. We instead focus on the residual magnetization of the final state when the bias field vanishes. We show on the basis of a phenomenologial argument as well as known critical exponents that the residual magnetization decays as the ramp speed $v$ as $v^{1/7}$. This scaling law is in marked contrast to the exponential scaling in the isolated spin model.

Numerical simulation of the time evolution in the dissipative spin model is not a trivial task due to the memory effect of the bath. Recently, the time-evolving matrix product operator (TEMPO) was proposed for this task\cite{strathearn2018,suzuki2019,oshiyama2022}. This method manages the memory effect by means of
a tensor network or a matrix product state in the time direction, allowing us to simulate the time evolution without the Born-Markov approximation. We apply this method to the study of the residual magnetization after a ramp.

The rest of the present paper is organized as follows. We introduce the isolated and the dissipative spin models in the next section. The Landau-Zener theory for the ramp of bias field is explained for the isolated spin model. Section \ref{sec3} is assigned to the scaling theory. We derive the scaling law of the residual magnetization after the half Landau-Zener ramp at the QPT of the dissipative model. In Sec. \ref{sec4}, after mentioning the TEMPO method briefly, we present results of our numerical simulation. The present paper is concluded in Sec. \ref{sec5}.

\section{Models}
\subsection{Landau-Zener ramp in the isolated spin model}

To begin with, we briefly review the Landau-Zener transition of an isolated spin. Let $\sigma^{\alpha}$
($\alpha = x$, $z$) be the Pauli spin operator, and $|\sigma\rangle$ ($\sigma = \uparrow$, $\downarrow$) be the eigenstate of $\sigma^z$; $\sigma^z|\uparrow\rangle = |\uparrow\rangle$,
$\sigma^z|\downarrow\rangle = - |\downarrow\rangle$. The Hamiltonian of an isolated spin with the Landau-Zener ramp is given by
\begin{equation}
 H_{\rm LZ}(t) = - \Delta\sigma^x + vt\sigma^z ,
\end{equation}
where $\Delta$ is the tunneling energy and $v$ is the speed of the ramp of the bias field. The time $t$ evolves from $t = -\infty$ to $+\infty$. We assume that the initial state at $t = -\infty$ is the ground state of $H_{\rm LZ}(-\infty)$, namely, $|\Psi(-\infty)\rangle = |\uparrow\rangle$ up to the phase factor.
The Schr\"odinger equation with this initial condition can be solved exactly. Using the parabolic cylinder function $D_p(z)$, the solution is written as \cite{zener1932,suzuki2013}
\begin{equation}
 |\Psi(t)\rangle = C_{\uparrow}(t)|\uparrow\rangle + C_{\downarrow}(t)|\downarrow\rangle ,
\end{equation}
with
\begin{eqnarray}
 C_{\uparrow}(t) &=& e^{-i\pi/4}e^{-\pi\Delta^2/8v}D_{-i\Delta^2/2v}\left(e^{-i 3\pi/4}\sqrt{2v}t\right),\\
 C_{\downarrow}(t) &=& \frac{\Delta}{\sqrt{2v}}e^{-\pi\Delta^2/8v}D_{-i\Delta^2/2v-1}\left(e^{-i 3\pi/4}\sqrt{2v}t\right).
\end{eqnarray}
The transition probability from the ground state $|\uparrow\rangle$ at $t=-\infty$
to the excited state $|\uparrow\rangle$ at $t=+\infty$ is given by
\begin{equation}
 P_{\uparrow\to\downarrow} = |C_{\uparrow}(+\infty)|^2
  = e^{-\pi\Delta^2/v} .
\label{eq:P-LZ}
\end{equation}
The magnetization at time $t$ is given by
\begin{equation}
 m(t) = \langle\Psi(t)|\sigma^z|\Psi(t)\rangle
  = |C_{\uparrow}(t)|^2 - |C_{\downarrow}(t)|^2
  = 2 |C_{\uparrow}(t)|^2 - 1
\label{eq:m-LZ}
\end{equation}
At $t=0$, formulas of the parabolic cylinder function \cite{abramowitz1972} leads to the following simple formula:
\begin{equation}
 m(0) = e^{-\pi\Delta^2/2v} .
\end{equation}

\subsection{Dissipative spin model}\label{sec2}

We consider the dissipative spin model (DSM) represented by the Hamiltonian:
\begin{equation}
 H = - \Delta\sigma^x - h\sigma^z
  + \sigma^z\sum_a\lambda_a(b_a^{\dagger} + b_a)
  + \sum_{a}\omega_a b_a^{\dagger}b_a ,
\label{eq:SB-model}
\end{equation}
where $b^{\dagger}$ and $b$ are the creation and annihilation operators of a boson with
mode $a$. $\Delta$ and $h$ denote the tunneling energy and the bias field, respectively. 
$\omega_a$ is the frequency of the harmonic oscillator of the mode $a$.
We choose the unit of $\hbar=1$.
The coupling constant $\lambda_a$ together with $\omega_a$ defines the
spectral density of the bath as
\begin{equation}
 J(\omega) = \sum_a\lambda_a^2\delta(\omega - \omega_a) ,
\end{equation}
which is assumed to be
\begin{equation}
 J(\omega) = \frac{\eta}{2}\omega_c\left(\frac{\omega}{\omega_c}\right)^s \exp\left(-\frac{\omega}{\omega_c}\right) ,
\end{equation}
where $\eta$ represents the coupling strength between the spin and the bath, and
$\omega_c$ is the cutoff frequency. The exponent $s$ determines the
character of the spectral density. In particular $s=1$ is referred to as the Ohmic bath, while
$0 < s < 1$ is the sub-Ohmic bath. We mainly focus on the sub-Ohmic bath in the present work.

Let us consider the partition function $Z = {\rm Tr}e^{-\beta H}$ with the inverse temperature $\beta$. Trotterizing the partition function \cite{trotter1959} and integrating the boson degrees of freedom, we obtain the path integral representation of the partition function as follows \cite{leggett1987,weiss2012}.
\begin{equation}
 Z = {\rm Tr}_{\sigma(\tau)}\exp\bigl(-\mathcal{S}[\sigma(\tau)]\bigr) ,
\label{eq:Z}
\end{equation}
where $\sigma(\tau)$ is an Ising-spin variable at an imaginary time $\tau$, and the action $\mathcal{S}[\sigma(\tau)]$ is written as
\begin{equation}
 \mathcal{S}[\sigma(\tau)] = \mathcal{S}_0[\sigma(\tau)]
  + \mathcal{S}_{\rm int}[\sigma(\tau)] ,
\end{equation}
with $\mathcal{S}_0[\sigma(\tau)]$ representing the free action for the isolated spin and
\begin{equation}
 \mathcal{S}_{\rm int} = - \int_0^{\beta}d\tau\int_0^{\tau}d\tau'
  \sigma(\tau)\mathcal{K}_{\beta}(\tau - \tau')\sigma(\tau') 
\end{equation}
representing the action for the spin-boson coupling. The kernel is given by
\begin{equation}
 \mathcal{K}_{\beta}(\tau) = \int_0^{\infty}d\omega J(\omega)\frac{\cosh(\beta\omega/2 - \tau\omega)}{\sinh\beta\omega/2}e^{-\omega/\omega_c} ,
\end{equation}
which for $\beta\to\infty$ reduces to
\begin{equation}
 \mathcal{K}_{\infty}(\tau) = \frac{\eta}{2}\omega_c^{1-s}
  \frac{1}{\left(\tau + \frac{1}{\omega_c}\right)^{s + 1}} .
\end{equation}
Therefore, $\mathcal{K}_{\infty}(\tau)$ behaves as $\tau^{-1-s}$ for $\tau\to\infty$. Note that a smaller $s$ implies a longer range. The ground state of DSM is attributed to the classical one-dimensional Ising model with a long-range interaction.

\section{Scaling theory}\label{sec3}

\begin{figure}[t]
 \begin{center}
  \includegraphics[width=5cm]{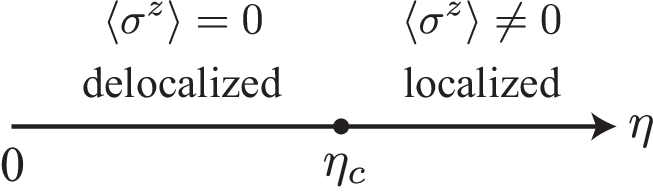}
 \end{center}
\caption{Schematic phase diagram for the ground state of DSM for a fixed $s\leq 1$.
}
\label{fig:phase_diagram_1}
\end{figure}

Assuming the absence of the bias field, $h=0$, DSM with $\eta = 0$ reduces to the isolated spin model only with the tunneling energy, while for $\eta/\Delta\to\infty$ the Hamiltonian is commutable with $\sigma^z$. The ground state is then delocalized for $\eta = 0$ while localized for $\eta/\Delta\to\infty$. A QPT at a critical point $\eta_c$ separates a delocalized phase with $\langle\sigma^z\rangle=0$ and a localized phase with $\langle\sigma^z\rangle\neq 0$ for $s\leq 1$ \cite{spohn1985,costi1999,bulla2003,winter2009}, see Fig.~\ref{fig:phase_diagram_1}. The magnetization $\langle\sigma^z\rangle$ serves as the order parameter in this transition.

The property of QPT in DSM is specified by the strength and the range of the interaction kernel $\mathcal{K}_{\infty}(\tau)$ parametrized by $\eta$ and $s$, respectively. The short-range interaction coming from $\mathcal{S}_0[\sigma(\tau)]$ is irrelevant to the critical property. Let us Define the effective Hamiltonian $\mathcal{H}$ by $\mathcal{S}_{\rm int} = \eta\mathcal{H}$. Then $\eta$ can be seen as the inverse temperature in the partition function given by Eq. (\ref{eq:Z}), and $s$ governs the spatial dimensionality of the effective Hamiltonian.  Thus, a set of a smaller $\eta$ and a larger $s$ is in favor of a delocalized phase, while a larger $\eta$ and a smaller $s$ is apt to yield a localized phase. It is known that the Kosterlitz-Thouless quantum transition separates a localized phase from a delocalized phase when $s = 1$. As for $0 < s < 1$, the numerical renormalization group and the quantum Monte-Carlo have revealed the existence of a critical line $\eta = \eta_c(s)$ of the second-order QPT in the $s$-$\eta$ plane\cite{bulla2003,winter2009}. Furthermore, the critical exponents of the spontaneous magnetization $\langle\sigma^z\rangle$, the magnetization under the bias field, the correlation length, and the dynamical exponent have been identified as $\beta = \frac{1}{2}$, $\delta = 3$, $\nu = \frac{1}{s}$, and $z = 1$ for $0 < s \leq \frac{1}{2}$\cite{winter2009,vojta2012}.

Let us focus on the scaling relation in the presence of the bias field $h$. The magnetization $m = \langle\sigma^z\rangle$ at the critical coupling strength $\eta_c(s)$ for a fixed $s$ ($0 < s < 1$)
scales as $m\sim h^{1/\delta}$. This implies the scaling of the free energy density
\begin{equation}
 f(\eta_c(s), h) \sim |h|^{1 + 1/\delta} ,
\label{eq:f-scaling1}
\end{equation}
so as to make $m = \frac{\partial f}{\partial h}\sim h^{1/\delta}$. We define here the correlation-length exponent $\tilde{\nu}$ at $\eta = \eta_c(s)$ and $h\neq 0$, giving rise to the scaling of the correlation length
$\xi(\eta_c(s), h) \sim |h|^{-\tilde{\nu}}$. The free energy density satisfies
\begin{equation}
 f(\eta_c(s), h) \sim \xi(\eta_c(s), h)^{-d} \sim |h|^{d\tilde{\nu}} .
\label{eq:f-scaling2}
\end{equation}
The relaxation time $t_r(\eta_c(s), h)$ diverges as 
\begin{equation}
 t_r(\eta_c(s), h)\sim \xi(\eta_c(s), h)^z \sim h^{-z\tilde{\nu}}
\label{eq:relaxation_time}
\end{equation}
with $|h|\to 0$. 
Equations (\ref{eq:f-scaling1}) and (\ref{eq:f-scaling2}) immediately lead to
\begin{equation}
 \tilde{\nu} = 1 + \frac{1}{\delta} .
\end{equation}
Applying $d = 1$ and the known numbers $\delta = 3$ for $0 < s < \frac{1}{2}$, we obtain
\begin{equation}
 \tilde{\nu} = \frac{4}{3} .
\end{equation}
This number together with $z = 1$ is used to derive the scaling of the
residual magnetization after the half Landau-Zener ramp.

\subsection{Scaling of the residual magnetization}\label{sec3.1}
Let us consider a ramp of the bias field by the schedule
\begin{equation}
 h(t) = -vt
\label{eq:h-schedule}
\end{equation}
with $t$ evolving from $t= -\infty$ to 0, so that $h(t)$ is ramped from $h=+\infty$
to $h= 0$ with the speed $v$. We call this the half Landau-Zener ramp. We assume that the full system is initialized at its ground state at $t = -\infty$. In the beginning, the full system evolves with keeping the ground state. On approaching the QPT, however, due to the growth of the relaxation time, the full system deviates from the ground state. At the end, when the bias field vanishes, the system acquires a finite residual magnetization. 

In order to clarify the relation between the residual magnetization and the ramp speed, we resort to the pulse-impluse approximation as follows. The full system keeps the ground state until the instantaneous relaxation time exceeds the remaining time to reach the critical point and it is frozen thereafter. The time when the full system is frozen is determined by the equality between the relaxation time $t_r(\eta_c(s), h(t))$ and the remaining time $|t|$:
$t_r(\eta_c(s), h(\hat{t})) = |\hat{t}|$. This equation together with Eqs. (\ref{eq:relaxation_time}) and (\ref{eq:h-schedule}) yields
\begin{equation}
 |\hat{t}| \sim v^{-z\tilde{\nu}/(z\tilde{\nu} + 1)} 
\end{equation}
up to a the nonuniversal factor. The bias field $\hat{h}$ at time $\hat{t}$ is
given by
\begin{equation}
 \hat{h} = h(\hat{t}) = v|\hat{t}| \sim v^{1/(z\tilde{\nu} + 1)} .
\end{equation}
Finally, the residual magnetization $m_{\rm res}$, defined by the magnetization at $h = \hat{h}$, is obtained as
\begin{equation}
 m_{\rm res} \sim \hat{h}^{1/\delta} \sim v^{1/\delta(z\tilde{\nu} + 1)} .
\end{equation}
Applying $\delta = 3$, $\tilde{\nu} = \frac{4}{3}$, and $z = 1$, this yields
for $0 < s \leq \frac{1}{2}$
\begin{equation}
 m_{\rm res} \sim v^{1/7} .
\label{eq:m-v}
\end{equation}

\section{Numerical simulation}\label{sec4}

\subsection{Method}

Since the QPT of our interest occurs due to the competition between the quantum tunneling and the coupling to a bath, these two terms must be treated equally. Hence the perturbative Born-Markov approximation is not a suitable approach to our purpose. We instead employ the TEMPO method, which we explain briefly here. For detail of this method, see refs. \cite{strathearn2018,suzuki2019}. 

\begin{figure}[t]
 \begin{center}
  \includegraphics[width=8cm]{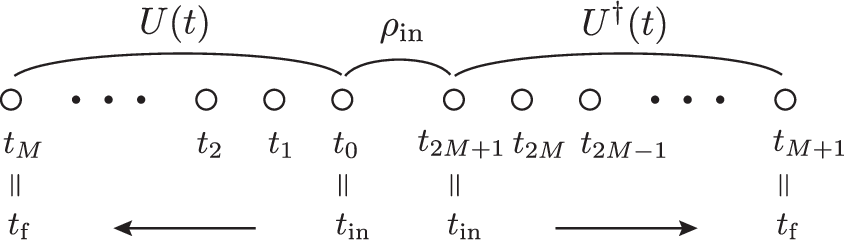}
 \end{center}
\caption{The definition of the discrete time. The arrows depict the direction of time evolution. Remark that the direction of time in $U^{\dagger}(t)$ is opposite from $U(t)$.
}
\label{fig:discrete_time}
\end{figure}

We consider the time-dependent density operator of the full system $\rho(t) = U(t)\rho_{\rm in}U^{\dagger}(t)$, where $U(t)$ is the unitary time-evolution operator from $t_{\rm in}$ to $t_{\rm f}$ and $\rho_{\rm in}$ is the density operator at the initial time $t_{\rm in}$. We assume that $\rho_{\rm in}$ is given by the direct product of the spin state and the boson state. Due to the presence of spin-boson coupling, the exact computation of the time-evolving density matrix is limited to short times. We here Trotterize $U(t)$ and integrate over the boson degrees of freedom \cite{makarov1994}. The resulting reduced density matrix at $t = t_{\rm f}$ in the eigenbasis of $\sigma^z$ is written as
\begin{equation}
 \langle\sigma_M|\rho_{\rm S}(t_{\rm f})|\sigma_{M+1}\rangle
  = \sum_{\sigma_0,\cdots,\sigma_{M-1}}\sum_{\sigma_{M+2},\cdots,\sigma_{2M+1}}
  e^{\mathcal{S}_0[\sigma_l] + \mathcal{S}_{\rm int}[\sigma_l]}
  \langle\sigma_0|\rho_{\rm S}(t_{\rm in})|\sigma_{2M+1}\rangle ,
\label{eq:DM-1}
\end{equation}
where $M$ denotes the Trotter number and we define the discrete time $t_l$ as
\begin{equation}
 t_l = \left\{\begin{array}{ll}
	t_{\rm in} + l\mathit{\Delta} t & (0\leq l\leq M) \\
	t_{\rm in} + (2M+ 1 - l)\mathit{\Delta} t & (M+1\leq l\leq 2M+1)\end{array}\right.
\end{equation}
with $\mathit{\Delta} t = \frac{t_{\rm f} - t_{\rm in}}{M}$. Figure \ref{fig:discrete_time} illustrates the definition of the discrete time. $\sigma_l$ denotes 
the spin variable at time $t_l$. $\rho_{\rm S}(t_{\rm in})$ is the spin state at $t = t_{\rm in}$. The free action $\mathcal{S}_0[\sigma_l]$ comes from the isolated spin, while $\mathcal{S}_{\rm int}[\sigma_l]$ is the action of the spin-boson coupling given by
\begin{equation}
 \mathcal{S}_{\rm int}[\sigma_l] = \mathit{\Delta} t^2\sum_{0\leq l<m\leq 2M+1}
  K(t_m - t_l)\sigma_l\sigma_m ,
\end{equation}
with the kernel function defined by
\begin{equation}
 K_{\beta}(t) = \int_0^{\infty} d\omega J(\omega)\frac{\cosh(\beta\omega/2 - i\omega t)}{\sinh\beta\omega/2} .
\end{equation}
This kernel function with $\beta\to\infty$ decays as $t^{-1-s}$ for $t\to\infty$.
For the purpose to reduce the cost of computation, we make a restriction to the range of $K(t)$ by introducing a cutoff time $t_c$ in such a way that $K(t) = 0$ for $t > t_c$.

Let us now assign the composite variable $S_l$ to the couple of variables $(\sigma_l, \sigma_{2M+1-l})$ ($l=0,1,\cdots,M$). Then, the action can be written as
\begin{equation}
 A_{S_0\cdots S_M} = e^{\mathcal{S}_0[\sigma_l] + \mathcal{S}_{\rm int}[\sigma_l]}
  \langle\sigma_0|\rho_{\rm S}(t_{\rm in})|\sigma_{2M+1}\rangle ,
\end{equation}
and seen as a tensor with $(M+1)$ indices. In TEMPO, this tensor is arranged into the form of a matrix product. Thus, the reduced density matrix (\ref{eq:DM-1}) is given in the form of
\begin{equation}
 \langle\sigma_M|\rho_{\rm S}(t_{\rm f})|\sigma_{M+1}\rangle
 = \sum_{S_0,\cdots,S_M}\sum_{p_1,\cdots, p_M}
 u_{p_1}^{S_0}u_{p_1,p_2}^{S_1}\cdots u_{p_{M-1},p_M}^{S_{M-1}}\psi_{p_M}^{S_M} .
\end{equation}
The matrix $u$ is obtained through the singular value decomposition. Its matrix
dimension is restricted by a so-called bond dimension $D_b$, so as to avoid the exponential increase of computational cost with repressing the truncation error.

The present method involves three approximations, Trotterization, the interaction cutoff in $K(t)$, and the bond dimension, which are controlled by $\mathit{\Delta} t$, $t_c$, and $D_b$, respectively. In our simulation, we fix $\mathit{\Delta} t = 0.0125/\Delta$ and $D_b = 128$, for which we have confirmed a good convergence. Regarding $t_c$, we set $t_c = 1/\Delta$ for $s = 0.5$ and $t_c=2/\Delta$ for $s= 0.3$. The reason for this choice is explained as follows. As mentioned above, the range of the interaction in the time direction is longer for smaller $s$. This implies that a larger $t_c$ is necessary for a smaller $s$. However, the computational cost increases with increasing $t_c$. Our choice is the consequence to have the largest $t_c$ under the requirement to keep an acceptable computational cost.

In our simulation, the bias field is changed from $h_0 = 20\Delta$ to $-20\Delta$ for the full Landau-Zener ramp and from $h_0 = 20\Delta$ to $0$ for the half Landau-Zener ramp, following the schedule (\ref{eq:h-schedule}) and the time $t$ moving from $t = - \frac{20\Delta}{v}$ to $\frac{20\Delta}{v}$ in the full Landau-Zener ramp and from $-\frac{20\Delta}{v}$ to 0 in the half Landau-Zener ramp. We set the initial state at the direct product of the ground state of the spin at $h=h_0$ and that of the bath. Since $h_0$ is sufficiently large compared to both the tunneling energy and the coupling strength to the bath, this initial state serves as a good approximation to the ground state of the full system.

\subsection{Results}

\begin{figure}[t]
 \begin{center}
 \includegraphics[width = 13cm]{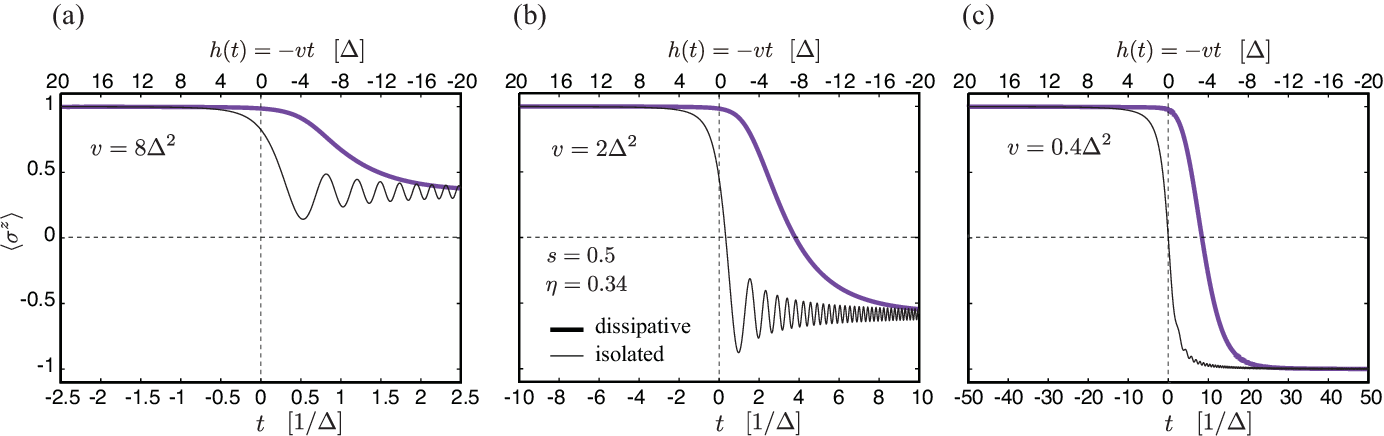}
 \end{center}
\caption{Time evolution of the magnetization during the ramp of the bias field following
the schedule (\ref{eq:h-schedule}). The horizontal axis is the time $t$
($-20\Delta/v \leq t \leq 20\Delta/v$), and the scale on the upper horizontal line denotes the bias field $h = -vt$. We show the results for the ramp speeds (a) $v = 8\Delta^2$, (b) $2\Delta^2$, and (c) $0.4\Delta^2$. 
The thick solid lines show the results in the presence of the spin-boson
coupling with $s = 0.5$ and $\eta = 0.34$, 
while thin lines are obtained by the isolated-spin model (\ref{eq:m-LZ}).
We set $t_c = 1/\Delta$ in TEMPO simulation. In case of the isolated spin, the magnetization decreases rapidly as the bias is vanishing, and converges with damped oscillation as the bias is decreased further in the negative side. In contrast, the magnetization of DSM remains at the unity until the bias vanishes and monotonically decreases when the bias is diminished further. Two curves of the isolated spin model and DSM for a fixed $v$ tends to converge to the same number for $t\to\infty$ $(h(t)\to -\infty)$.}
\label{fig:mz-h}
\end{figure}

\begin{figure}[t]
 \begin{center}
  \includegraphics[width=8cm]{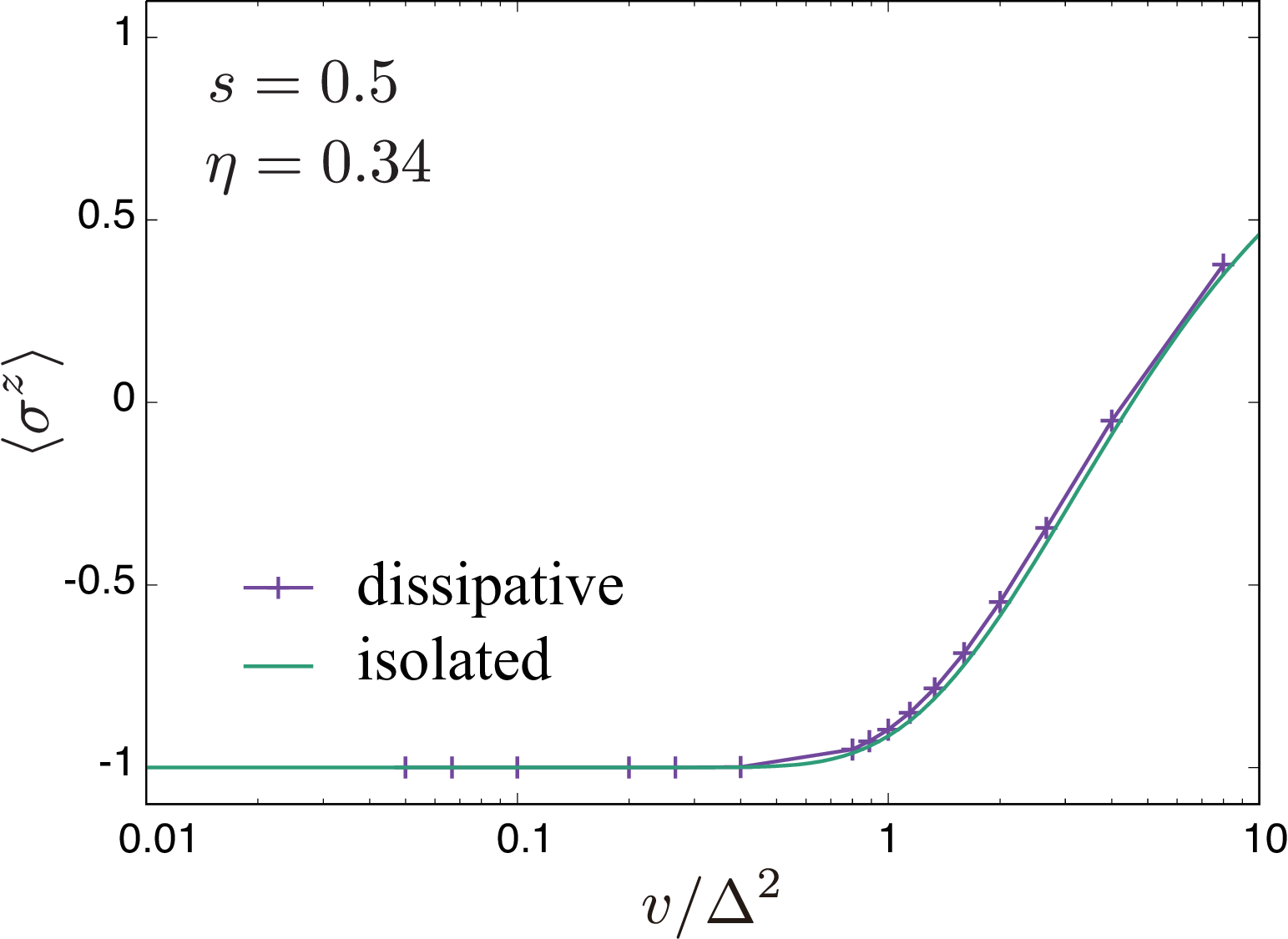}
 \end{center}
\caption{Magnetization after the full ramp of the bias field as a function the ramp speed $v$. In DSM, the bias field is flipped from $h=h_0 = 20\Delta$ to $-20\Delta$ with time period $40\Delta/v$. The curve for the isolated spin shows $m(\infty) = -1 + 2 e^{-\pi\Delta^2/v}$, obtained from Eqs. (\ref{eq:m-LZ}) and (\ref{eq:P-LZ}). There is a clear agreement between the isolated-spin model and DSM. As far as the full ramp is concerned, the bath does not influence the state after the ramp.}
\label{fig:mz-v-full}
\end{figure}

We first compare the dynamics of the full Landau-Zener ramp in the isolated-spin and the dissipative models. For the dissipative model, we focus on $s = 0.5$ and fix $\eta = 0.34$ which is slightly smaller than $\eta_c(s = 0.5)\approx 0.4$ \cite{winter2009}. Figure \ref{fig:mz-h} shows the time dependence of the magnetization during the full Landau-Zener ramp. While the magnetization drops when the bias field vanishes in the isolated-spin model, the descent happens much later when the bias is considerably negative in the dissipative model. In the both models, the magnetization is likely to converge to the same $v$-depending number for $t\to\infty$. Figure \ref{fig:mz-v-full} shows the $v$-dependence of the magnetization for long $t$. The data of the TEMPO simulation were obtained at $t = 20/v$, and compared to the analytic result of the isolated-spin model for $t\to\infty$ which is given by Eqs. (\ref{eq:m-LZ}) and (\ref{eq:P-LZ}).
The numerical result for the dissipative model agrees with the curve for the isolated-spin model. This is consistent with an exact theory in Ref.~\cite{wubs2006} that the boson bath does not influence the transition probability $P_{\uparrow\to\downarrow}$ given by (\ref{eq:P-LZ}) for the evolution from $t=-\infty$ to $+\infty$, demonstrating the correctness of the present numerical simulation.

\begin{figure}[t]
 \begin{center}
  \includegraphics[width=8cm]{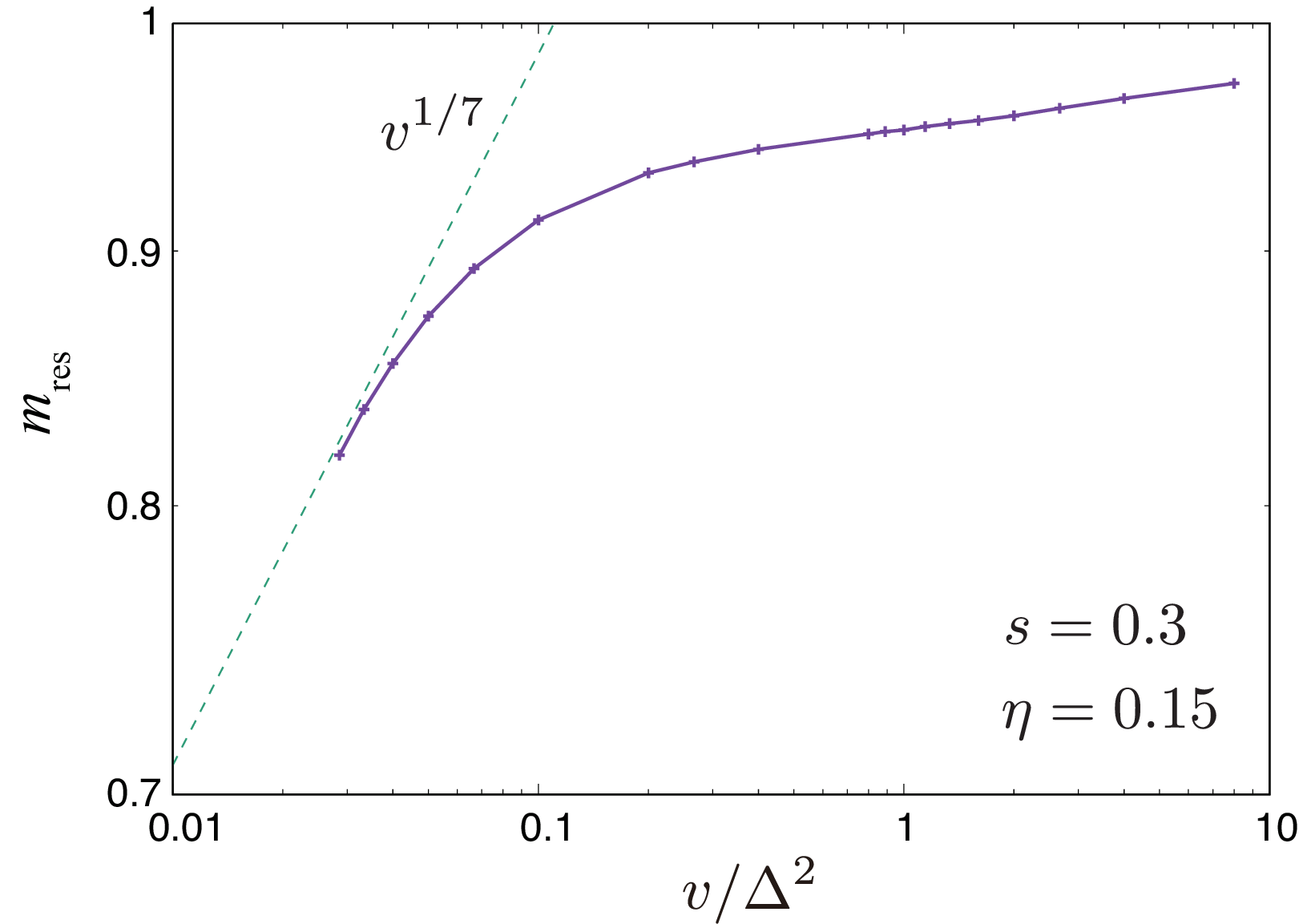}
 \end{center}
\caption{Residual magnetization after the half Landau-Zener ramp toward the quantum critical point ($\eta_c\sim 0.15$ for $s=0.3$ \cite{winter2009}) as a function of the ramp speed $v$. The dashed line indicates a function proportinal to $v^{1/7}$ and is a guide to the eye. The residual magnetization is likely to follow $v^{1/7}$ for small $v$. The bias field is ramped from $h_0 = 20\Delta$ to $0$. We set $t_c = 2/\Delta$.}
\label{fig:mz-v-half}
\end{figure}

Figure \ref{fig:mz-v-half} shows the scaling of the residual magnetization after the half Landau-Zener ramp toward the quantum critical point in the dissipative model. The bath has the sub-Ohmic spectral density with $s=0.3$ and the system-bath coupling is fixed at the critical point $\eta_c(s=0.3) \approx 0.15$, which is extracted from Ref. \cite{winter2009}, so that the half Landau-Zener ramp terminates at the quantum critical point.  The residual magnetization is likely to follow $v^{1/7}$ for small $v$. Due to the computational cost, the data are not available for $v$ slower than $0.028$. Although one cannot rule out other possibilities, the numerical result shown in Fig. \ref{fig:mz-v-half} is consistent so far with the theoretical prediction of Eq. (\ref{eq:m-v}).


\section{Conclusion}\label{sec5}

We studied the magnetization of a single spin coupled to a sub-Ohmic boson bath in the presence of a ramp of the bias field. It has been known that, when the bias field is ramped from $+\infty$ to $-\infty$, the magnetization after the ramp follows the Landau-Zener formula irrespective of the coupling strength and the spectral property of the bath. However, this is not true when a ramp stops intermediately. The present work focused on the situation where the bias field is ramped toward a quantum phase transition. We derived that the residual magnetization after the ramp scales as $v^{1/7}$ with the ramp speed $v$ for the sub-Ohmic bath with $0 < s \leq \frac{1}{2}$, by means of the pulse-impulse approximation and with known critical exponents. This scaling was supported to some degree, for $s = 0.3$ in particular, by our numerical experiments using TEMPO.

Our numerical data have not been sufficient so far to verify the predicted scaling of the residual magnetization. To obtain clearer evidence of the power law scaling, a more systematic numerical study for different cutoff times would be necessary. It would be also significant to investigate the bath with finite temperature. Since the range of interaction in the time direction is shortened due to the finite-temperature bath, computational cost would be reduced and a more direct indication of scaling law would be detected. This line of study is left as a future work.

The use of quantum simulation is another approach. The physical system realizing the DSM has an advantage over the classical computer in the sense that the long-range interaction in time is naturally involved. The most promising system is superconducting flux qubits hosting a quantum annealer. While it is difficult to control the coupling strength between the spin and a bath, one can tune the tunneling energy and ramp the bias field. Therefore, the half Landau-Zener ramp can be realized in the system of quantum annealer. The experimental study along this line would be another future work.

\bmhead{Acknowledgements}
The present paper is devoted to Professor Amit Dutta, who conducted extensive research on out-of-equilibrium phenomena in the vicinity of quantum phase transitions, the subject of this work. He excelled in employing dimensional arguments, which play an essential role in the theory of quantum phase transitions, and the author gained valuable insights from discussions with him. His passing has left the author with a profound sense of loss. The author expresses deepest gratitude to Professor Amit Dutta.
The author acknowledges Hiroki Oshiyama for providing the codes for TEMPO simulations.
The present work was supported by JSPS KAKENHI grant No. 22K03455.

\section*{Declarations}

Numerical data will be made available on reasonable request.

%
%


\bibliography{halfLZbib}

\end{document}